\documentclass[12pt]{JHEP3}
\usepackage[fleqn]{amsmath}
\usepackage{bbm,bm}

\let\br\-

\def\S{S}
\def\F{\phi}
\def\PS{\hat{\S}}
\def\PF{\F}
\def\p{{(+)}}

\def\m{{(-)}}
\def\ppm{{(\pm)}}
\def\pmp{{(\mp)}}

\def\+{{+\!\!\!+}}
\def\-{=} 
\def\pp{{\mbox{\tiny${}_{\stackrel\+ =}$}}}

\def\eps{\epsilon}

\def\smallhalf{{\textstyle \frac{1}{2}}}
\def\ph#1{\phantom{#1}}
\def\phn{{\ph{0}}}

\def\d{{\rm d}}
\def\i{{\rm i}}
\def\half{{\frac{1}{2}}}
\def\id{{\mathbbm{1}}}

\def\+{{+\!\!\!+}}
\def\-{=} 
\def\ph#1{\phantom{#1}}
\def\phn{{\phantom{0}}}
\def\d{{\rm d}}

\def\gcg#1{{\cal #1}}
\ifx\bm\undefined
  \def\gen#1{{\bf #1}}
\else
  \def\gen#1{{\bm #1}} 
\fi
\def\genJ{\gen{J}}
\def\genR{\gen{R}}
\def\genG{\gen{G}}
\def\genB{\gen{B}}
\def\genH{\gen{H}}
\def\genU{\gen{U}}

\def\genGamma{{\gen{\Gamma}}}
\def\genLambda{{\gen{\Lambda}}}
\ifx\mathbbm\undefined
  \def\genop#1{{\mathbb{#1}}}
\else
  \def\genop#1{{\mathbbm{#1}}}
\fi
\ifx\bm\undefined
  \def\genpartial{{\partial\hspace*{-1.2ex}\partial}}
  \def\gennabla{{\nabla\hspace*{-2ex}\nabla}}
  \def\bm\bf
\else 
  \def\genpartial{{\gen{\partial}}}
  \def\gennabla{{\gen{\nabla}}}
\fi

\def\matrix#1#2{\left(\begin{array}{#1}#2\end{array}\right)}
\def\genMatrix#1{\matrix{ccc}{#1}}
\def\gcgMatrix#1{\matrix{cc}{#1}}
\def\D{{\overrightarrow{D}}}
\def\Dt{{\overleftarrow{D}}}
\def\diag{{\rm diag}}

\ifx\mathbbm\undefined
  \def\chiral#1{{\mathbb{#1}}}
\else
  \def\chiral#1{{\mathbbm{#1}}}
\fi
\def\achiral#1{{\bar{\chiral{#1}}}}
\def\i{{\rm i}}

\title{First-order supersymmetric sigma models and target space geometry
}

\author{
  \parbox{14cm}{
    Andreas Bredthauer\footnotemark[1], 
    Ulf Lindstr{\"o}m\footnotemark[1],\footnotemark[4] 
    Jonas Persson\footnotemark[1]}\\
  \parbox[t]{6.5cm}{
   \footnotemark[1]\,\,\parbox[t]{6cm}{
     Department of Theoretical Physics\\
     Uppsala University\\
     Box 803, SE-751\,08 Uppsala\\
     Sweden
   }
  }
  \parbox[t]{6.5cm}{
   \footnotemark[4]\,\,\parbox[t]{6cm}{
     Helsinki Institute of Physics\\
     P.O. Box 64\\
     FIN-000\,14 University of Helsinki\\
     Finland\\
   }
  }
  \ \\   
  {\tt Andreas.Bredthauer, Ulf.Lindstrom, Jonas.Persson@teorfys.uu.se}
}

\abstract{
We study the conditions under which $N=(1,1)$ generalized sigma models support
an extension to $N=(2,2)$. The enhanced supersymmetry is related to the target
space complex geometry. Concentrating on a simple situation, related to Poisson
sigma models we develop a language that may help us analyze more complicated
models in the future. In particular, we uncover a geometrical framework which
contains generalized complex geometry as a special case. 
}

\preprint{
  hep-th/0508228\\
  UUITP-15/05\\
  HIP-2005-35/TH\\
}


\begin{document}
\section{Introduction}

Supersymmetry has a number of interesting relations to geometry. The analogue
of Minkowski space is superspace, whose geometry is nontrivial even in the
`flat' case \cite{Gates:1983nr}. Curved superspace is the setting for
supergravity and has a wealth of interesting geometrical aspects
\cite{Gates:1983nr, Buchbinder:1998qv,Wess:1992cp}. Superembeddings in curved
superspace constrains the geometry very stringently and  in many cases even
determines the dynamics of the embedded super $p$-branes \cite{Howe:2004ib}.
Extended supersymmetry is covariantly described in various extended superspaces
with auxiliary degrees of freedom \cite{Karlhede:1984vr, Gonzalez-Rey:1997qh,
Galperin:2001uw}. The target space of supersymmetric nonlinear sigma models,
finally, has to be of a certain type depending on the dimension and on the
number of supersymmetries.  It is this latter situation which concerns us in
this paper, more precicely the geometry of twodimensional $N=(2,2)$
supersymmetric nonlinear sigma models with an antisymmetric $B$-field.

In a classic paper \cite{Gates:nk} it was shown that the target space of such a
sigma model has to be  bi-hermitean, i.e.\ there are two complex structures
preserving the metric and they are covariantly constant with respect to
connections whose torsions are $\pm \d B$. Recently this geometry has been
reinterpreted in terms of a generalized complex geometry, which arose in the
context of generalized Calabi-Yau manifolds with $B$-field fluxes 
\cite{Hitchin:2004ut}.  In \cite{Gualtieri:2004} many aspects of this geometry
are investigated and described. In particular, it is shown that a subclass
called generalized K\"ahler geometry precicely describes the bi-hermitean
geometry.

A natural question to ask is then how generalized K\"{a}hler geometry can be
directly realized in a sigma model.  Since generalized complex geometry is
defined on the sum of the tangent and cotangent bundles, $TM\oplus T^*M$, and
the usual  sigma model is defined only on $TM$, the first task is to find an
appropriate extension of the sigma model to include fields on $T^*M$.  This was
done in \cite{Lindstrom:2004eh} where auxiliary spinorial $T^*M$-fields were
introduced in the $N=(1,1)$ model and the conditions for  non-manifest
$N=(2,2)$ supersymmetry investigated under certain assumptions. This
investigation was repeated in \cite{Bergamin:2004sk}, for the case when the
metric is absent. Relaxing these assumptions and limiting the study mainly to
extending $N=(1,0)$ to ${N=(2,0)}$, a direct relation to generalized complex
geometry was found in in most cases \cite{Lindstrom:2004iw}.  However, in that
investigation it seemed that the geometry in the $N=(2,0)$ case might be even
more general, although the study was incomplete. 

To further investigate the geometry, a manifest $N=(2,2)$ model in terms of
left and right (anti-)chiral superfields \cite{Buscher:1987uw} was reduced to
$N=(1,1)$ superfields and the generalized complex structures identified in
\cite{Lindstrom:2004hi}. An interesting aspect of this model is that the
reduction automatically provides the auxiliary spinorial $N=(1,1)$ fields.

In a separate line of investigation \cite{Zucchini:2004ta,Zucchini:2005rh}, it
has been shown that generalized complex geometry bears a close relation to the
Batalin-Vilkovisky (BV) treatment of the Poisson sigma model, or more precisely
to the Hitchin sigma model. Namely, the generalized complex geometry implies
that the BV-master equation is satisfied.  Also in this case the implication
seems to go only in one direction. Generalized complex geometry has also
appeared in the sigma model context, e.g.\ in a hamiltonian discussion
\cite{Zabzine:2005qf} and for topological strings \cite{Kapustin:2004gv}.

The reason that the investigation of the conditions for $N=(2,0)$ supersymmetry
(and for $N=(2,2)$ supersymmetry) was not carried out \cite{Lindstrom:2004iw}
was mainly the technical complications of having to find solutions to a large
number of algebraic and differential constraints. In the present paper we show
how an appropriate field-redefinition can be used to put the sigma model action
in a form where invariance under supersymmetry restricts many of the tensors in
the supersymmetry transformations of the fields to vanish.  This allows us to
completely determine the target space geometry, at least for the case of
vanishing metric, i.e.\ with only a $B$-field  present. In doing this we
unravel a target space structure where the natural objects live on $TM\oplus
(T^*M_{+}\oplus T^*M_{-})$, i.e.\ the geometry involves two copies of the
cotangent bundle rather than one.  Correspondingly all the fundamental
geometric objects such as almost complex structures, metric and connections
have a natural formulation in terms of $3d\times 3d$ matrices. In some respects
this structure resembles the bi-hermitean geometry of the second order action
(auxiliary fields removed) more than the generalized K\"ahler geometry. In
particular, the Courant integrability condition of the generalized complex
geometry is replaced by covariantly constancy of the matrix-valued almost
complex structures. Now, one of the nice features of generalized complex
geometry is that it naturally puts the so-called $b$-transform on the same
footing as the diffeomorphisms since they are both automorphisms of the
Courant-bracket. It is thus gratifying that we find that the  $b$-transform can
be extended to act on our matrix-objects, and that this extended $b$-transform
is indeed a gauge transformation of our basic bundle which preserves the
covariantly constancy condition. Finally, under certain conditions the
$3d\times 3d$ matrices collapse to $2d\times 2d$ matrices recovering
generalized complex geometry.  In other words, the latter is contained in the
structure we have found.

The paper is organized as follows: After a short recapitulation of the basic
facts about $N=(2,2)$ supersymmetric sigma models in section 2, we turn towards
a toy model which we extend to a first order formalism in section 3. For this
model, we give a huge family of solutions for the additional supersymmetry that
all close off-shell. Section 4 is devoted to the development of a proper
language that collects the results in a way similar to the notion of
generalized complex geometry. Based on these results, we discuss in section 5
how to find more general solutions. In section 6, we show how this relates to the
geometry of $N=(2,2)$ symplectic sigma models in a way that extends the
$b$-transformation.  In section 7 we speculate about the role of
manifest $N=(2,2)$ supersymmetry before ending with a short discussion and open
questions in section 8.

\section{$N=(2,2)$ sigma models, preliminaries}
The action for a $N=(1,1)$ supersymmetric non-linear sigma model under the
presence of a background metric $G_{\mu\nu}$ and an antisymmetric field
$B_{\mu\nu}$
\begin{gather}
  S=\int \d^2\xi \d^2\theta D_+ \F^\mu E_{\mu\nu}(\F) D_- \F^\nu
  \label{e:S=DFEDF}
\end{gather}
possesses $N=(2,2)$ supersymmetry \cite{Gates:nk} provided that the target space
geometry is bi-hermitian. Here, $D_\pm$ are the spinorial derivatives,
$D_\pm^2=\i\partial_\pp$, and $E_{\mu\nu}=G_{\mu\nu}+B_{\mu\nu}$. The
additional, non-manifest supersymmetry is given by
\begin{gather}
  \delta\F^\mu = \eps^+ J^{\p\mu}_\nu D_+\F^\nu + \eps^- J^{\m\mu}_\nu D_-\F^\nu
  \label{e:delF}
\end{gather}
where $J^\ppm$ are complex structures. The metric is hermitian with respect to
both of them and the complex structures are covariantly constant, i.e.
\begin{gather}
 \begin{aligned}
    &J^{\ppm2} = -\id \hspace*{5cm}&& N(J^\ppm) = 0 \\
    &G_{\mu\nu} = J^{\ppm\kappa}_\mu G_{\kappa\lambda}J^{\ppm\lambda}_\nu &&
    \nabla^\ppm_\rho J^{\ppm\mu}_\nu = 0.
 \end{aligned}
 \label{e:Jcond}
\end{gather}
Here, $N(J^\ppm)$ is the Nijenhuis torsion for $J^\ppm$, 
\begin{gather}
  N(J^\ppm)^\mu_{\alpha\beta} = 
	  J^{\ppm\mu}_\rho J^{\ppm\rho}_{[\beta\alpha]}
	- J^{\ppm\rho}_{[\alpha}J^{\ppm\mu}_{\beta]\rho}.
\end{gather}
The covariant derivatives $\nabla^\ppm$ are given by the connections
\begin{gather}
  \Gamma^{\ppm\alpha}_{\beta\gamma} = \Gamma^\alpha_{\beta\gamma} \pm
    T^\alpha_{\beta\gamma}
\end{gather}
where $\Gamma^\alpha_{\beta\gamma}$ is the metric connection and 
$T^\alpha_{\beta\gamma} = \half H_{\beta\gamma\kappa}G^{\kappa\alpha}$ is the
torsion. This implies that $H=\d B$ is related to the complex structures in a
certain way.

The above conditions ensure that the additional supersymmetry commutes with the 
first manifest supersymmetry and that its algebra closes on-shell. 
Off-shell closure is achieved provided that the two complex structures commute,
\begin{gather}
  {[}J^\p,J^\m] = 0.
\end{gather}
This and \eqref{e:Jcond} imply that the Magri-Morosi concomitant
\cite{Magri:1984, Yano:1968ab}
\begin{gather}
  M(J^\p, J^\m)^\mu_{\alpha\beta} = 
	  J^{\p\mu}_{\alpha\rho}J^{\m\rho}_\beta
	- J^{\m\mu}_{\beta\rho}J^{\p\rho}_\alpha
	+ J^{\p\mu}_\rho J^{\m\rho}_{\alpha\beta}
	- J^{\m\mu}_\rho J^{\p\rho}_{\beta\alpha}
\end{gather} 
vanishes and that both complex structures and the product structure 
$\pi = J^\p J^\m$ are integrable and simultaneously diagonalizable.

While in the previous discussion the metric $G_{\mu\nu}$ played a crucial role,
we now repeat the analysis in the case of an antisymmetric background field
$B_{\mu\nu}$ only, i.e.\ we set $E_{\mu\nu} \equiv B_{\mu\nu}$ in the action
\eqref{e:S=DFEDF} and obtain
\begin{gather}
  S_B=\int \d^2\xi \d^2\theta D_+ \F^\mu B_{\mu\nu}(\F) D_- \F^\nu.
  \label{e:S=DFBDF}
\end{gather}
Requiring off-shell supersymmetry, we learn that the set of constraints on the
transformations \eqref{e:delF} reduces to
\begin{gather}
  \begin{aligned}
    &J^{\ppm2} = -1 
    \hspace*{2cm} &&N(J^\ppm) = 0 
    \hspace*{2cm} &&H = 0 \\
    & {[}J^\p,J^\m] = 0 
    &&M(J^\p,J^\m) = 0 .
  \end{aligned}
\end{gather}
Thus, the target-space geometry is bicomplex. The condition $H=0$ implies that
the model is topological. This is a perfect toy model for our purpose, as we see
it as a first step towards understanding more general sigma models with extended
supersymmetry.

\section{Auxiliary fields and supersymmetry algebra} \label{s:solution}
First order sigma model actions have recently come into the focus of research
due to their relation to generalized complex geometry on the target manifold.
While it is straightforward but lengthy to work out the on-shell supersymmetry
transformations \cite{Lindstrom:2004eh}, off-shell supersymmetry is still not
really understood in geometrical terms, partly due to the lack of notation.
Several attempts were made to identify those models that admit or require
generalized complex geometry
\cite{Bergamin:2004sk,Lindstrom:2004iw,Zucchini:2004ta,Zucchini:2005rh,
        Zabzine:2005qf,Zabzine:2004dp,Lyakhovich:2002kc}. 
Here, we follow a different approach to investigate the question of off-shell
supersymmetry. We focus on the action \eqref{e:S=DFBDF} and introduce
spinorial auxiliary fields $\S_{\pm}$ on $T^*M$. They are combined into an
auxiliary term added to the action
\begin{gather}
  S = \int \d^2\xi \d^2\theta
    \Big[ \S_{+\mu}\Pi^{\mu\nu}\S_{-\nu}
         +D_+\F^\mu B_{\mu\nu} D_-\F^\nu \Big] \label{e:action} .
\end{gather}
To keep things simple, we assume that $\Pi$ is a Poisson tensor of full rank,
i.e.\ it is symplectic and hence satisfies the Jacobi identity
$\Pi^{[\alpha\beta}{}_\rho \Pi^{\rho|\gamma]}=0$.

By dimensional arguments, see e.g.\ \cite{Lindstrom:2004eh}, the most general 
form of the second supersymmetry is given by
\begin{align}
 \begin{split}
  \delta ^{\ppm}\F^\mu=&
     \eps^{\pm}\left(D_{\pm}\F^\nu
     J^{\ppm\mu}_{\nu}-\S_{\pm\nu}P^{\ppm\mu\nu}\right)
     \\
  \delta^{\ppm} \S_{\pm\mu}=&
    \eps^{\pm}\left(D_\pm^2\F^\nu
    L^{\ppm}_{\mu\nu}-D_{\pm}\S_{\pm\nu}K^{\ppm \nu}_{\mu}
    +\S_{\pm\nu}\S_{\pm\sigma}N^{\ppm\nu\sigma}_{\mu}\right.\cr
  &\left. \qquad \qquad +D_{\pm}\F^\nu D_{\pm}\F^\rho
    M^{\ppm}_{\mu\nu\rho}+D_{\pm}\F^\nu \S_{\pm\sigma}
    Q^{\ppm\sigma}_{\mu\nu}\right) \\
  \delta^{\ppm} \S_{\mp\mu}=&
    \eps^{\pm}\left(D_{\pm}\S_{\mp\nu}R^{\ppm\nu}_{\mu}+
    D_{\mp}\S_{\pm\nu}Z^{\ppm\nu}_{\mu}+D_{\pm}D_{\mp}\F^\nu
      T^{\ppm}_{\mu\nu}\right.
    \cr
  &\left. \qquad \qquad +\S_{\pm\rho}D_{\mp}\F^\nu U^{\ppm\rho}_{\mu\nu}
    +D_{\pm}\F^\nu \S_{\mp\rho} V^{\ppm\rho}_{\mu\nu}\right.\cr
  &\left. \qquad \qquad +D_{\pm}\F^\nu D_{\mp}\F^\rho X^{\ppm}_{\mu\nu\rho}
    +\S_{\pm\nu}\S_{\mp\rho}Y^{\ppm\nu\rho}_{\mu}\right).
  \end{split}
  \label{e:delFS+-}
\end{align}
The action \eqref{e:action} is invariant under these transformations provided
that
\begin{align}
  \Pi^{\mu\alpha} R^\beta_\alpha &= -K^{\mu}_\nu \Pi^{\nu\beta} &
  \Pi^{(\alpha|\rho}Z^{\beta)}_\rho &= 0 &
  L_{\alpha\beta} &= 0 &
  T_{\alpha\beta} &= 0 \label{e:invariance-action}
\end{align} 
and that a set of differential equations hold. One of these is
\begin{math}
  H = 0. 
\end{math}
For the time being, we make
the assumption that $P^\p$ and $P^\m$ are invertible. It turns out that things
simplify drastically under this assumption. Indeed, already the commutators of the
second supersymmetry with itself provide 113 conditions to be satisfied. We
comment on the situation for more general $P^\ppm$ in section \ref{app:A}.
Off-shell closure of the additional supersymmetry algebra is guaranteed if
$J^{\ppm}$ are commuting complex structures that are covariantly constant with
respect to certain torsionfree
connections $\Gamma^{(J^\ppm)\mu}_{\nu\rho}$
\begin{align}
 \begin{aligned}
  &J^{\ppm2} = -1 &&&
  &{[}J^\p, J^\m] = 0 &&& 
  &\nabla^{(J^\ppm)} J^{\ppm} = 0 &&&
 \end{aligned} \label{e:J2=-1}
\end{align}
The transformations
\eqref{e:delFS+-} are determined by the composite tensors:
\begin{align}
 \begin{split}
  K^{\ppm\beta}_\alpha &= -P^{\ppm}_{\alpha\mu}J^{\ppm\mu}_\nu P^{\ppm\nu\beta}\\ 
  R^{\ppm\beta}_\alpha &= -\Pi_{\alpha\mu}K^{\ppm\mu}_\nu \Pi^{\nu\beta} \\
  L^{\ppm}_{\alpha\beta} &= 0\\
  T^{\ppm}_{\alpha\beta} &= 0\\
  Z^{\ppm\alpha}_\beta &= 
	-P^{\pmp}_{\beta\kappa}P^{\ppm\kappa\lambda}R^{\pmp\alpha}_\lambda
     	+P^{\pmp}_{\beta\kappa}J^{\pmp\kappa}_\lambda P^{\ppm\lambda\alpha}
 \end{split}\label{e:lower-order-tensors} 
\end{align}
\begin{align}
 \begin{split}
  M^{\ppm}_{\mu\nu\rho} &= 0 \\
  X^{\ppm}_{\mu\nu\rho} &= 0 \\
  Q^{\ppm\rho}_{\mu\nu} &=
        \Gamma^{(K^{\ppm})\rho}_{\beta\mu}J^{\ppm\beta}_\nu
        +\Gamma^{(K^{\ppm})\rho}_{\nu\kappa}K^{\ppm\kappa}_\mu  \\ 
  V^{\ppm\beta}_{\mu\gamma} &= 
       \Gamma^{(R^{\ppm})\beta}_{\rho\mu}J^{{\ppm}\rho}_\gamma
       - \Gamma^{(R^{\ppm})\beta}_{\gamma\rho}R^{\ppm\rho}_\mu  \\
  U^{\ppm\alpha}_{\beta\gamma} &=
     \Gamma^{(K^{\ppm})\alpha}_{\gamma\kappa}Z^{\ppm\kappa}_\beta \\ 
  N^{\ppm[\alpha\beta]}_\gamma &=
      \Gamma^{(K^{\ppm})[\alpha}_{\kappa\gamma}P^{\ppm\kappa|\beta]} \\ 
  Y^{\ppm\alpha\beta}_\gamma &= -\Gamma^{(R^{\ppm})\beta}_{\rho\gamma}P^{\ppm\rho\alpha},
 \end{split} \label{e:solutionhigherorder}
\end{align}
where
\begin{align}
 \begin{split}
  \Gamma^{(K^{\ppm})\sigma}_{\rho\lambda}
    &= \big[ P^{\ppm}_{\lambda\mu\rho}
	 - P^{\ppm}_{\lambda\nu} \Gamma^{(J^{\ppm})\nu}_{\rho\mu}  
	 \big] P^{\ppm\mu,\sigma} \\
  \Gamma^{(R^{\ppm})\sigma}_{\rho\lambda}
    &= \big[ \Pi_{\lambda\mu,\rho} 
	- \Pi_{\lambda\nu} \Gamma^{(K^{\ppm})\nu}_{\rho\mu}  
	\big] \Pi^{\mu\sigma}.
 \end{split}
\end{align}

The second rank tensors are `covariantly constant' according to
\begin{align}
 \begin{split}
   \nabla_\rho J^{\ppm\alpha}_\beta &\equiv
        J^{\ppm\alpha}_{\beta,\rho} 
		- \Gamma^{(J^{\ppm})\nu}_{\rho\beta}J^{\ppm\alpha}_\nu
                + \Gamma^{(J^{\ppm})\alpha}_{\rho\nu}J^{\ppm\nu}_\beta = 0 \\
   \nabla_\rho K^{\ppm\alpha}_\beta &\equiv
        K^{\ppm\alpha}_{\beta,\rho} 
		- \Gamma^{(K^{\ppm})\nu}_{\rho\beta}K^{\ppm\alpha}_\nu
                + \Gamma^{(K^{\ppm})\alpha}_{\rho\nu}K^{\ppm\nu}_\beta = 0 \\
   \nabla_\rho R^{\ppm\alpha}_\beta &\equiv
        R^{\ppm\alpha}_{\beta,\rho} 
		- \Gamma^{(R^{\ppm})\nu}_{\rho\beta}R^{\ppm\alpha}_\nu
                + \Gamma^{(R^{\ppm})\alpha}_{\rho\nu}R^{\ppm\nu}_\beta = 0 \\
   \nabla_\rho P^{\ppm\alpha\beta} &\equiv P^{\ppm\alpha\beta}{}_{,\rho}
      		+ P^{\ppm\alpha\nu} \Gamma^{(K^{\ppm})\beta}_{\rho\nu}
       		+ \Gamma^{(J^{\ppm})\alpha}_{\rho\nu} P^{\ppm\nu\beta} = 0 \\
   \nabla_\rho Z^{\ppm\alpha}_\beta &\equiv
    Z^{\ppm\alpha}_{\beta,\rho} 
		- \Gamma^{(R^{\ppm})\nu}_{\rho\beta} Z^{\ppm\alpha}_\nu
      		+ \Gamma^{(K^{\ppm})\alpha}_{\rho\nu}Z^{\ppm\nu}_\beta =0 .
 \end{split}
 \label{e:cov-const}
\end{align}
The connections are related as
\begin{align}
   \Gamma^{(J^\m)} = \Gamma^{(J^\p)} &&
   \Gamma^{(R^{\ppm})} = \Gamma^{(K^{\pmp})} \label{e:connections-rel}.
\end{align}
The corresponding Riemann tensors 
\begin{math}
  R^{(\cdot)\kappa}{}_{\lambda\mu\nu} = \Gamma^{(\cdot)\kappa}_{[\nu|\lambda,|\mu]}
  +\Gamma^{(\cdot)\rho}_{[\nu|\lambda}\Gamma^{(\cdot)\kappa}_{\mu]\rho}
\end{math}
vanish:
\begin{gather}
  R^{(R^\ppm)}=R^{(K^\ppm)}=R^{(J^\ppm)}=0 \label{e:Riemann-vanish}.
\end{gather}
From the non-derivative parts of the algebra, one constraint remains:
\begin{gather}
	  R^{\ppm\rho}_\mu Z^{\ppm\alpha}_\rho 
	+ Z^{\ppm\rho}_\mu K^{\ppm\alpha}_\rho = 0.
\end{gather}
We observe that, except for being covariantly constant, there is no constraint on
$P^\ppm$. Equations \eqref{e:lower-order-tensors} imply
\begin{align}
  Z^{\ppm\alpha}_\beta &= P^\pmp_{\beta\kappa}{[}J^\pmp, 
    P^\m\Pi^{-1}P^{\p t}]^{\kappa\alpha} \label{e:Z=PJPiP}.
\end{align}
The relation \eqref{e:Z=PJPiP} shows that it is possible, at least in
certain situations, to choose $P^\ppm$ in such a way that both $Z^\ppm$ vanish.
This requires both complex structures to commute with
\begin{math}
\omega^{\alpha\beta}\equiv(P^\m\Pi^{-1}P^{\p t})^{\alpha\beta}
 = -(P^\p\Pi^{-1}P^{\m t})^{\beta\alpha} 
\end{math}.
In other words, $\omega$ has to be antihermitian with respect to both complex
structures. If, on the other hand, $\omega$ is antisymmetric and in additions
satisfies the Jacobi identity then we may identify its inverse with the
two-form of a symplectic manifold.  Clearly, $G^{\ppm} = J^{\ppm}\omega$ are then
candidates for effective metrics. One such example is the case
$P^{\m\alpha\beta}=P^{\p\alpha\beta}$. It follows that $R^{\ppm} = -K^{\ppm}$,
$[K^\p,K^\m] = 0$ and $\Pi$ is antihermitian with respect to $K^{\ppm}$.
However, this alternative is only possible if $\Pi$ is covariantly constant.
\begin{align}
  \nabla^{(K)}_\rho \Pi^{\alpha\beta} =
    \Pi^{\alpha\beta}{}_{,\rho} + 
       \Gamma^{(K)[\alpha}_{\rho\kappa}\Pi^{\kappa|\beta]} = 0, 
\end{align}
where $\Gamma^{(K)}\equiv \Gamma^{(K^\p)}=\Gamma^{(K^\m)}$.

This covers the discussion of the second supersymmetry transformations under the
assumptions \eqref{e:J2=-1} for the particular model we study. Equation
\eqref{e:connections-rel} is sufficient for off-shell closure. It 
might not be necessary though we find this quite unlikely due to the
way \eqref{e:connections-rel} contributes to the solution.

\section{Almost complex structures on $TM\oplus(T^*M_+\oplus T^*M_-)$}
In the previous section we found that the complete data identifying the
solution is encoded in the objects $B$, $\Pi$, $J^{\ppm}$, $P^{\ppm}$ and
$\Gamma^{(J)}$. We want a formulation as closely related as possible to
generalized complex geometry \cite{Hitchin:2004ut,Gualtieri:2004} and shall try
to find a role for the components of \eqref{e:delFS+-} in that context. We start
with a recapitulation of the notion of generalized complex geometry. 

An almost complex structure is a linear map $J:TM\rightarrow TM$ that squares to 
$-1$. If we define projection operators $\pi_\pm =\smallhalf (1\pm\i J)$, then
$J$ is integrable if 
\begin{gather}
  \pi_\mp{[}\pi_\pm X,\pi_\pm Y] = 0 \label{e:proj-integrability}
\end{gather}
for any $X,Y \in TM$, where $[\cdot,\cdot]$ is the Lie bracket on $TM$. Hitchin
\cite{Hitchin:2004ut} proposed and later Gualtieri \cite{Gualtieri:2004}
investigated a generalization of this notion, where $TM$ is replaced by
$TM\oplus T^*M$ and the Lie bracket is replaced by the so-called Courant
bracket. A generalized complex structure is defined as a map
$\gcg{J}:TM\oplus T^*M \rightarrow TM\oplus T^*M$, such that $\gcg{J}^2=-1$ and
it leaves the natural symmetric inner product
\begin{align}
  \langle X+\xi, Y+\eta \rangle = \smallhalf(i_X \eta + i_Y \xi) && 
	X+\xi, Y+\eta \in TM\oplus T^*M
\end{align}
invariant. In a coordinate basis $(\partial_\mu, \d x^\mu)$, the metric
\begin{gather}
  \gcg{I} = \genMatrix{0&1\\1&0}
\end{gather}
is hermitian with respect to $\gcg{J}$. Furthermore, the $+\i$ eigenbundle of
$\gcg{J}$ is closed under the Courant bracket \cite{Courant:1990}, which is
defined as 
\begin{gather}
  [X+\xi,Y+\eta]_C=[X,Y]+L_X\eta-L_Y\xi - \smallhalf\d(i_X\eta - i_Y\xi).
\end{gather}
This bracket allows to define Courant integrability as a straightforward
generalization of \eqref{e:proj-integrability}. In a coordinate basis,
generalized complex structures can be written in terms of $2d\times 2d$ matrices
\begin{gather}
  \gcg{J} = \gcgMatrix{
	J&P\\L&K }.
\end{gather}
An important feature of the Courant bracket is the existence of non-trivial
automorphisms defined by closed two-forms $b\in \Omega^2_{\rm closed}(M)$.
Consequently, given a generalized complex structure $\gcg{J}$, we can define
a new such structure by the $b$-transformation
\begin{align}
  \gcg{J}_b = \gcg{U}\gcg{J}\gcg{U}^{-1} &&
  \gcg{U} = \gcgMatrix{\phn 1&\phn 0\\-b&\phn 1}. \label{e:gcg-b-trans}
\end{align}
The automorphism of the Courant bracket guarantees this structure to be
integrable. For a detailed discussion, we refer to the original works
\cite{Hitchin:2004ut, Gualtieri:2004}.

In \cite{Lindstrom:2004iw}, the authors constructed examples of sigma models
admitting generalized complex geometry in the target space. Mainly as a
curiosity, they found that the algebraic conditions for closure of the algebra
could be combined into a single $3d\times 3d$ matrix squaring to $-1$. This
object seems like a natural extension of the concept of generalized complex
structures. Here, we elaborate this idea in detail and use it as a basis for the
description of the target space geometry. We thus combine the tensors into two
$3d\times 3d$ matrices 
\begin{align}
  {\genJ}^{\p} = \genMatrix{%
             \ph{-}J^\p&-P^\p&\ph{-}0 \\
             -L^\p&\ph{-}K^\p&\ph{-}0 \\
             \ph{-}T^\p&-Z^\p&\ph{-}R^\p
             }&&
  {\genJ}^\m = \genMatrix{%
             \ph{-}J^\m&\ph{-}0&-P^\m \\
             \ph{-}T^\m&\ph{-}R^\m&-Z^\m \\
             -L^\m&\ph{-}0&\ph{-}K^\m
             } \label{e:genJ+-}.
\end{align}
The components of these matrices are the linear maps 
\begin{gather}
 \begin{aligned}
  &J^\p: TM\rightarrow TM && P^\p: T^*M_+ \rightarrow TM \\
  &L^\p: TM\rightarrow T^*M_+ && K^\p: T^*M_+ \rightarrow T^*M_+ \\
  &T^\p: TM\rightarrow T^*M_- && Z^\p: T^*M_+ \rightarrow T^*M_- &&
   R^\p: T^*M_-\rightarrow T^*M_-.
 \end{aligned}
\end{gather}
The components of $\genJ^\m$ are defined analogously. Here, $T^*M_+$ and
$T^*M_-$ are two copies of the cotangent bundle. They are associated with the
two Grassmann directions on the worldsheet. Thus, $\genJ^\ppm$ map the bundle
$E=TM\oplus (T^*M_+ \oplus T^*M_-)$ onto itself. Guided by the action
\eqref{e:action} we introduce a (degenerate) symmetric inner product on $E$,
an equivalent to the metric for the ordinary sigma model:
\begin{gather}
  {\genG} = \genG^t = \half \genMatrix{
             \ph{-}0&\ph{-}0&\ph{-}0 \\
             \ph{-}0&\ph{-}0&\ph{-}\Pi \\
             \ph{-}0&\ph{-}\Pi^t&\ph{-}0
             }. 
\end{gather}
We note that $\genG$ is degenerate because we set $E_{(\mu\nu)}=0$ in
\eqref{e:S=DFBDF} and that $\genG$ is antisymmetric in the fermionic components.
The algebraic conditions arising from the invariance of the action, eqns.\
\eqref{e:invariance-action}, and the non-differential part of the algebra
\eqref{e:lower-order-tensors} can be written in a compact way:
\begin{align}
  {\genJ}^{\ppm t}{\genG} {\genJ}^{\ppm} = {\genG}&&
  {\genJ}^{\ppm2} = - {\bf 1} &&
  {[}{\genJ}^\p,{\genJ}^\m] = 0. \label{e:solutioncompact}
\end{align}
This allows us to regard $\genJ^\ppm$ as (almost) complex structures on $E$.
Eqns.\ \eqref{e:cov-const} tell us that these structures are covariantly
constant, 
\begin{gather}
  \gennabla^{\ppm}{\genJ}^{\ppm} \equiv \genpartial {\genJ}^{\ppm}
     -\genJ^{\ppm}\cdot \genGamma^{\ppm}
     +{\genGamma}^{\ppm}\cdot {\genJ}^{\ppm} = 0 \label{e:nablacalJ}
\end{gather}
with respect to certain connection matrices
\begin{gather}
 \begin{split}
  {\genGamma}^{\p} = 
    {\diag}\left(\Gamma^{(J^{\p})},-\Gamma^{(K^{\p})},-\Gamma^{(R^{\p})}\right) \\
  {\genGamma}^{\m} = 
    {\diag}\left(\Gamma^{(J^{\m})},-\Gamma^{(R^{\m})},-\Gamma^{(K^{\m})}\right)
\end{split}
\end{gather}
and a partial derivative $\genpartial = {\bf 1}\partial$. Equation
\eqref{e:connections-rel} translates into
\begin{align}
  \genGamma \equiv \genGamma^\p = \genGamma^\m .
\end{align}
The components of $\genGamma$ are torsionfree, $\genGamma^t=\genGamma$, where
the transposition is acting on the two lower indices, and its Riemann tensor is
\begin{gather}
   {\genR} = {[}\gennabla,\gennabla] = 
     {\bf d}\genGamma - {\genGamma}\circ{\genGamma}
   \label{e:genR}
\end{gather}
where ${\bf d} = \gen{1}\d$ is the generalized exterior derivative. According to
\eqref{e:Riemann-vanish}, this matrix vanishes:
\begin{gather}
  {\genR} = 0.
\end{gather}
In K{\"a}hler geometry, the Nijenhuis torsion and the Levi-Civita
connection are related by
\begin{gather}
  N(J)(X,Y)=(\nabla_{JX}J)Y-(\nabla_{JY}J)X+(\nabla_X J)JY - (\nabla_Y J)JX,
\end{gather}
with $X,Y\in TM$.  Clearly, if $J$ is covariantly constant with respect to the
Levi-Civita connection, then $N(J)=0$. The generalization to a matrix-valued
Nijenhuis torsion $\gen{N}(\genJ^\ppm)$ would make use of $\gennabla$ and
$\genGamma$ and hence vanishes if $\gennabla\genJ^\ppm=0$. Thus,
\eqref{e:nablacalJ} is an integrability condition ensuring the integrability of
$J^\ppm,~K^\ppm$ and $R^\ppm$.

We find that the above description completely covers closure of the
supersymmetry algebra and most of the conditions that arise from the invariance
of the action. In fact, the only condition left is $H=0$.
We define an antisymmetric tensor by
\begin{gather}
  {\genB} = \half\genMatrix{%
                    \ph{-}2B&\ph{-}0&\ph{-}0\\
                    \ph{-}0&\ph{-}0&\ph{-}\Pi\\
                    \ph{-}0&-\Pi^t&\ph{-}0
            	    } 
\end{gather}
and define its field strength in the usual way,
\begin{gather}
  {\genH} = {\rm\bf d}\genB = \half\genMatrix{%
                    \ph{-}2H_B&\ph{-}0&\ph{-}0\\
                    \ph{-}0&\ph{-}0&\ph{-}\Pi H_\Pi \Pi \\
                    \ph{-}0&\ph{-}\Pi H_\Pi \Pi&\ph{-}0
            	    }. 
\end{gather}
Here, $H_\Pi=\d (\Pi^{-1})$  which vanishes in our case, since $\Pi$ is symplectic. 
With this, we have
\begin{gather}
  {\genH} = 0.
\end{gather}

There are actually four different possibilities for choosing the two 
almost complex structure matrices describing one and the same situation. They
are obtained from \eqref{e:genJ+-} by acting on
$\genJ^\ppm$ with $\gen{C}^\ppm=\diag(1,\mp 1,\pm 1)$ and
$\gen{S}=\gen{C}^\p\gen{C}^\m$:
\begin{gather}
  \begin{aligned}
    \genJ^\ppm_1 &= \genJ^\ppm \hspace*{2cm}&  
    \genJ^\ppm_2 &= \gen{C}^\p\genJ^\ppm_1\gen{C}^\p \\
    \genJ^\ppm_3 &= \gen{S}\genJ^\ppm_1\gen{S} &
    \genJ^\ppm_4 &= \gen{C}^\m\genJ^\ppm_1\gen{C}^\m.
   \end{aligned}
\end{gather}
The covariant derivative is changed accordingly, e.g.\
\begin{align}
    \genpartial_2 = \gen{C}^\p\genpartial &&
    \genGamma_2 = \gen{C}^\p\genGamma. 
\end{align}
This symmetry is reminicent of the discrete symmetries of the first order sigma model
action discussed in, e.g.\ \cite{Lindstrom:2004eh}. The whole discussion may
equally well be formulated in terms of any of these choices.

This completes the discussion of the model \eqref{e:action} in this language. 
However, it is worth noticing that the geometry of the ordinary second order sigma
model \eqref{e:S=DFEDF} is embedded in this framework in a natural way. It corresponds to
\begin{align} 
  \genG=\diag(G,0,0)&& \genB=\diag(B,0,0).
\end{align}
Of course, then $\genGamma^\p$ and $\genGamma^\m$ are no longer related in the same
way, since $B$ generates torsion in the tangent space directions.

\section{Towards a more general solution} \label{app:A}
One of the main ingredients of the solution given in section \ref{s:solution} is
the invertibility of $P^\ppm$. This assumption was made because the conditions
for the supersymmetry algebra to close simplified drastically. This helped
us to introduce the compact notation in the previous section. However, the
spacetime geometry turned out to be completely empty, since there is neither a
metric nor a three-form field strength. Here, we elaborate the 
case where $P^\ppm$ may have degeneracies. This implies that the tangent bundle
complex structures $J^\ppm$ are no longer related to the cotangent bundle 
ones $K^\ppm,~R^\ppm$ in a unique way. The non-differential conditions for
invariance of the action and closure of the algebra are still ensured by 
\eqref{e:solutioncompact}
\begin{align}
  \genJ^{\ppm2}=-1&& {[}\genJ^\p,\genJ^\m] = 0 && \genJ^{\ppm t}
\gen{G}\genJ^\ppm =\gen{G}.
\end{align}
We observe that the higher order tensors of the solution
\eqref{e:solutionhigherorder} do not depend on $\Gamma^{(J^\ppm)}$ but rather on
the connections for $K^\ppm$ and $R^\ppm$. This allows us to go beyond flat
space in the following way:
To stick as close as possible to the solution given in the previous
sections, we start with the assumption that there are two connections
$\Gamma^{(R^\ppm)}$ such that $R^{\ppm}$ are two covariantly constant complex
structures. With this, the solution on the two copies of the cotangent bundle
$T^*M_+\oplus T^*M_-$ remains the same as before, since
\begin{align}
  \Gamma^{(K)\epsilon}_{\rho\mu} = - \Pi^{\epsilon\nu}\big[
  \Gamma^{(R)\sigma}_{\rho\nu}\Pi_{\sigma\mu} + \Pi_{\nu\mu,\rho}\big].
\end{align}
and since closure of the algebra requires $R^{(R^\ppm)} = R^{(K^\ppm)} = 0$. In
order for the higher order tensors to remain defined as in 
\eqref{e:solutionhigherorder}, we need the further assumption that there exists 
$A^{\ppm\alpha}_{\rho\nu}$ such that
\begin{align}
   P^{\ppm\alpha\beta}{}_{,\rho}
      		+ P^{\ppm\alpha\nu} \Gamma^{(K^{\ppm})\beta}_{\rho\nu}
       		= A^{\ppm\alpha}_{\rho\nu} P^{\ppm\nu\beta}.
\end{align}
Together with the equation
\begin{align}
  \nabla_\sigma^{(K^\ppm)}\big[ J^{\ppm\mu}_\rho P^{\ppm \rho \alpha}
   +P^{\ppm\mu\rho}K^{\ppm\alpha}_\rho \big] = 0,
\end{align}
we read off the connection for the tangent bundle 
$\Gamma^{(J^\ppm)^\alpha}_{\rho\nu} = - A^\alpha_{\rho\nu}$ and learn that
$J^{\ppm}$ only has to be covariantly constant in the
directions where $P^{\ppm}$ is invertible. 
On ${\rm ker}(P)$, we do no longer get any differential conditions and thus,
locally, the tangent space geometry becomes bicomplex. This fits to the original 
second order sigma model with a $B$-field only where we obtained a bicomplex 
geometry and no differential conditions for $J^\ppm$. Especially for
$P^{\ppm\mu\nu}=0$, we recover this situation, as expected, since the supersymmetry 
transformation for $\phi^\mu$ decouples from the auxiliary fields $S_{\pm\mu}$.
Since $R^{(J^\ppm)}$ now in general is non-vanishing, we obtain a more involved 
geometry of the tangent bundle, while the cotangent bundle does not
carry any additional geometric structure.


\section{Symplectic sigma model and $B$-transformation}
The $N=(2,2)$ supersymmetric symplectic sigma model action 
\cite{Lindstrom:2004eh,Schaller:1994es}
\begin{gather}
  S_{SSM} = \int \d^2\xi \d^2\theta \Big[
    \PS_{(+\mu}D_{-)}\PF^\mu + \PS_{+\mu}\Pi^{\mu\nu}\PS_{-\nu} \Big]
\end{gather}
is obtained from \eqref{e:action} by the transformation
\begin{gather}
  \S_{\pm\mu}\rightarrow \PS_{\pm\mu} = \S_{\pm\mu} - \Pi_{\mu\nu}D_\pm \F^\nu
  \label{e:StoS_PSM}
\end{gather}
and by identifying $B_{\mu\nu}\equiv \Pi_{\mu\nu}$. To be a bit more general,
however, we consider the action
\begin{gather}
  S_{SSM+B} = S_{SSM} + \int \d^2\xi \d^2\theta \Big[
    D_+\PF^\mu (B_{\mu\nu} - \Pi_{\mu\nu}) D_-\PF^{\nu} \Big].
  \label{e:S_PSM+B}
\end{gather}
We notice that if we take $\Pi_{\mu\nu}$ to be a globally defined two-form, this is
precisely the action used to discuss the WZW term in \cite{Lindstrom:2004iw}
with the metric set to zero. By rewriting the transformations \eqref{e:delFS+-}
in terms of $\PF$ and $\PS_\pm$, we obtain the contributions to the new tensors,
which we denote by a hat to distinguish them from the previous results. We omit
the $\ppm$ for a better legibility.
\begin{align}
 \begin{split}
  \hat J^\mu_\nu  &= J^\mu_\nu - P^{\mu\rho} \Pi_{\rho\nu} \\
  \hat P^{\mu\nu} &= P^{\mu\nu} \\
  \hat L_{\mu\nu} &= \Pi_{\mu\rho}\hat J^\rho_\nu - K_\mu^\rho \Pi_{\rho\nu} \\
  \hat K^\nu_\mu  &= K^\nu_\mu + \Pi_{\mu\rho} P^{\rho\nu}\\
  \hat T_{\mu\nu} &= (R^\rho_\mu - Z^\rho_\mu)\Pi_{\rho\nu} - \Pi_{\mu\rho}
                     \hat J^\rho_\nu \\
  \hat Z^\nu_\mu  &= Z^\nu_\mu - \Pi_{\mu\rho} P^{\rho\nu} \\
  \hat R^\nu_\mu  &= R^\nu_\mu
 \end{split} \label{e:gen-components}
\end{align}
\begin{align}
\begin{split}
  \hat N^{[\mu\nu]}_\rho &= N^{[\mu\nu]}_\rho \\
  \hat M_{\mu[\nu\rho]} &= \Pi_{\mu\kappa}\hat J^{\kappa}_{[\rho\nu]} 
     +\Pi_{\mu[\nu|,\kappa}\hat J^\kappa_{\rho]}
     +K^\kappa_\mu \Pi_{\kappa[\nu,\rho]}
     -N^{[\kappa\lambda]}_\mu \Pi_{\kappa\nu}\Pi_{\lambda\rho}
     -Q^\kappa_{\mu[\nu}\Pi_{\kappa|\rho]} \\
  \hat Q^\rho_{\mu\nu} &=Q^\rho_{\mu\nu}
     - \Pi_{\mu\kappa}P^{\kappa\rho}{}_\nu
     + N^{[\kappa\rho]}_\mu \Pi_{\kappa\nu}
     - \Pi_{\mu\nu,\kappa} P^{\kappa\rho}\\
  \hat U^\rho_{\mu\nu} &= U^\rho_{\mu\nu}
     + \Pi_{\mu\sigma}P^{\sigma\rho}{}_\nu
     + \Pi_{\mu\nu,\sigma} P^{\sigma\rho}
     + Y^{\rho\sigma}_\mu \Pi_{\sigma\nu}\\
  \hat V^\rho_{\mu\nu} &= V^\rho_{\mu\nu}
     + Y^{\rho\sigma}_\mu \Pi_{\sigma\nu}\\
  \hat X_{\mu\nu\rho} &= -\Pi_{\mu\sigma} \hat J^\sigma_{\nu\rho}
     - \Pi_{\mu\rho,\sigma}\hat J^\sigma_\nu
     + (R^\sigma_\mu - Z^\sigma_\mu) \Pi_{\sigma\nu,\rho} \cr
  &\hspace*{4cm}
     + U^\sigma_{\mu\rho} \Pi_{\sigma\nu}
     + V^\sigma_{\mu\nu} \Pi_{\sigma\rho}
     + Y^{\kappa\lambda}_\mu \Pi_{\kappa\nu}\Pi_{\lambda\rho}\\
  \hat Y^{\nu\rho}_\mu &= Y^{\nu\rho}_\mu.
\end{split}
\end{align}
The transformation of $\genJ^\ppm$ with components \eqref{e:gen-components} can
be written in a compact way: 
\begin{align}
  \hat \genJ^\ppm = {\genU}\genJ^\ppm {\genU}^{-1} &&
  {\genU} = \genMatrix{%
     \ph{-}1&\ph{-}0&\ph{-}0 \\
     -\Pi^{-1}&\ph{-}1&\ph{-}0 \\
     -\Pi^{-1}&\ph{-}0&\ph{-}1
     }. \label{e:J=UJU}
\end{align}
This implies that
\begin{align}
 \begin{split}
  \hat{\genG} &=  ({\genU}^{-1})^{t}{\genG}{\genU}^{-1} =
    \half\genMatrix{%
      \ph{-}0&\ph{-}1&-1\\
      \ph{-}1&\ph{-}0&\ph{-}\Pi\\
      -1&-\Pi&\ph{-}0
      } \\
  \hat{\genB} &=  ({\genU}^{-1})^{t}{\genB}{\genU}^{-1} =
    \half\genMatrix{%
      2(B-\Pi^{-1})&-1&-1\\
      \ph{-}1&\ph{-}0&\ph{-}\Pi\\
      \ph{-}1&\ph{-}\Pi&\ph{-}0
      }.
 \end{split} 
\end{align}
$\hat{\genG}$ is hermitian with respect to $\hat\genJ^{\ppm}$,
$\hat{\gen{H}} = 0$ and ${\genU}$ is unitary. If we regard
\eqref{e:StoS_PSM} as a gauge transformation, that is, an automorphism of the
bundle $E$, then $\genGamma$ transforms as a connection and \eqref{e:nablacalJ}
is invariant,
\begin{math}
 \hat{\gennabla}\hat{\genJ}^\ppm = \gen{U}\gennabla\genJ^\ppm \gen{U}^{-1}= 0.
\end{math} 
Equations \eqref{e:J=UJU} and \eqref{e:StoS_PSM} extend the $b$-transform
\eqref{e:gcg-b-trans} of generalized complex geometry to our formulation. Hence,
it is suggestive to regard \eqref{e:nablacalJ} as an integrability condition.
It is puzzling how to fit in \eqref{e:StoS_PSM} in a proper way. Obviously,
\begin{align}
  \hat{\genLambda} \neq {\genU}{\genLambda}&&{\genLambda} = (\F, \S_+, \S_-)^t
  \label{e:Lambda=ULambda}
\end{align}
due to the derivatives on $\F$. In generalized complex geometry, this problem
does not occur, since the fermionic derivative can be included in the definition
of $\genLambda$. Here, there are two of them, $D_\pm$, which complicates the
situation. To inspect this in more detail, we promote the matrices to operators
in the following way: 
\begin{gather}
\begin{aligned}
  \genop{U} &= \genMatrix{%
     \ph{-}1&\ph{-}0&\ph{-}0 \\
     -\Pi^{-1} D_+&\ph{-}1&\ph{-}0 \\
     -\Pi^{-1} D_-&\ph{-}0&\ph{-}1
     } \hspace*{1cm} &&
  \genop{G} &= \genMatrix{%
             \ph{-}0&\ph{-}0&\ph{-}0 \\
             \ph{-}0&\ph{-}0&\ph{-}\Pi \\
             \ph{-}0&\ph{-}\Pi^t&\ph{-}0
           } \\
  \genop{B} &= \genMatrix{%
                    {\Dt}_{(+} B{\D}_{-)}&\ph{-}0&\ph{-}0\\
                    \ph{-}0&\ph{-}0&\ph{-}\Pi\\
                    \ph{-}0&-\Pi^t&\ph{-}0
             	}.
\end{aligned}
\end{gather}
Even if B is antisymmetric, 
the inner product \begin{math}
  \langle\genLambda_1,\genLambda_2\rangle \equiv
  \genLambda_1^t\genop{B}\genLambda_2
\end{math}
is actually symmetric in $\genLambda_1,\genLambda_2$. Accordingly, the
almost complex structure matrices become
\begin{align}
 \begin{split}
  \genop{J}^\p &= 
	\genMatrix{%
		\ph{-}J D_+   &-P         &\ph{-}0 \\
		-L D_+^2      &\ph{-}K D_+&\ph{-}0 \\
		\ph{-}T D_+D_-&-Z D_-     &\ph{-}R D_+
		} \\
  \genop{C}^\p &= \gen{C}^\p = {\diag}(1, -1, 1).
 \end{split}
\end{align}
We introduce the following object:
\begin{align}
 \begin{split}
  \genop{Q}^\p &= (\genop{Q}^{\p\F},\genop{Q}^{\p\S_+},\genop{Q}^{\p\S_-})^t \\
  \genop{Q}^{\p\F}&= 0 \\
  \genop{Q}^{\p\S_+} &= \genMatrix{%
		\Dt_+ M \D_+ 	& \smallhalf\Dt_+ Q& \ph{-}0 \\
		\smallhalf Q\D_+& \ph{-}N	   & \ph{-}0 \\
		\ph{-}0		& \ph{-}0	   & \ph{-}0
		} \\
  \genop{Q}^{\p\S_-} &= \genMatrix{%
		\Dt_+ X \D_-	& \ph{-}0 	  & \smallhalf \Dt_+ V \\
		\ph{-}0		& \ph{-}0 	  & \smallhalf \Dt_- U \\
		\smallhalf V\D_+& \smallhalf U\D_-& Y
		} \\
 \end{split}
\end{align}
and $\genop{J}^\m$, $\genop{Q}^\m$, $\genop{C}^\m$ defined correspondingly.
With this notation the transformation is given by
\begin{align}
  \hat\genLambda = \genop{U} \genLambda &&
  \hat{\genop{G}} = (\genop{U}^{-1})^t \genop{G} (\genop{U}^{-1}) &&
  \hat{\genop{B}} = (\genop{U}^{-1})^t \genop{B} (\genop{U}^{-1}).
\end{align}
The action \eqref{e:action} can be written as
\begin{gather}
  S = \frac{1}{2}\int \d^2\xi \d^2\theta \, \genLambda^t \genop{E} \genLambda = 
  \frac{1}{2}\int \d^2\xi \d^2\theta \, \hat\genLambda^t \hat{\genop{E}}
     \hat\genLambda
\end{gather}
where $\genop{E} = \genop{G} + \genop{B}$.
The supersymmetry transformations become
\begin{gather}
  \delta^{\ppm}\genLambda = \eps^\pm \genop{C}^\ppm\genop{J}^{\ppm}\genLambda
    +\eps^\pm \genLambda^t \genop{Q}^{\ppm} \genLambda .
\end{gather}
Thus, the matrices $\gen{C}^\ppm$ arise here as well.  Closure of the algebra
reduces to the condition
\begin{gather}
  {[}\delta_1, \delta_2]\genLambda = 2\eps^+_1\eps^+_2 \partial_\+\genLambda
    +2\eps^-_1\eps^-_2 \partial_\-\genLambda.
\end{gather}
It is not difficult to check that this operator formulation works also for 
(ordinary) second order sigma models and in the context of generalized complex
geometry on $TM\oplus T^*M$.

\section{Manifest supersymmetry and left-/right-chiral superfields}
There are several ways to construct manifest $N=(2,2)$ sigma models by using 
constrained $N=(2,2)$ superfields. The different possibilities are chiral,
twisted chiral and left-/right-chiral ones together with their antichiral partners
\cite{ 	Gates:nk,
        Lindstrom:2004hi,
	Abou-Zeid:1999em,
	Sevrin:1996jr,
	Sevrin:1996jq}.
To understand how the latter, originally called semichiral fields, fit into our
description in terms of $N=(1,1)$ manifest supersymmetry, we start with the
simple toy model action
\begin{gather}
  S = -\int \d^2\xi \d^2\theta \d^2\bar\theta \Big[ 
    \chiral{X}{\achiral{Y}} - {\achiral{X}}\chiral{Y} \Big]
    = \int \d^2\xi \d^2\theta \d^2\bar\theta \Big[
      \chiral{X}^A B_{AB^\prime} \chiral{Y}^{B^\prime}\Big]
  \label{e:N=2.2action}
\end{gather}
$\chiral{X},\chiral{Y}$ are the left-chiral and right-antichiral superfields
\cite{Buscher:1987uw}:
\begin{gather}
 \begin{aligned}
  &\achiral{D}_+\chiral{X} = 0 \hspace*{3cm}&& \chiral{D}_-\chiral{Y} = 0 \\
  &D_\pm = \chiral{D}_\pm + \achiral{D}_\pm &&
  Q_\pm = \i(\chiral{D}_\pm - \achiral{D}_\pm)  \\
  &\varphi = \chiral{X}| && \Psi_- = Q_-\chiral{X}| \\
  &\chi = \chiral{Y}| && \Upsilon_+ = Q_+\chiral{Y}| .
 \end{aligned}
 \label{e:XYdef}
\end{gather}
The $p$ left chiral and $p^\prime$ right antichiral holomorphic coordinate
indices $a$ and $a^\prime$ and their antiholomorphic partners are conveniently
collectively denoted $A=a,\bar a$ and $A^\prime=a^\prime,\bar a^\prime$. Moreover, we
introduce $\alpha =A,A^\prime$.

After a redefinition of the fields, 
\begin{gather}
 \begin{split}
  (\Psi_-,\bar\Upsilon_+) \rightarrow (\hat\Psi_-,\hat{\bar\Upsilon}_+) = 
    (\Psi_-,\bar\Upsilon_+) - \i (D_-\varphi, D_+\bar\chi) \\
  ({\bar\Psi}_-,{\Upsilon}_+) \rightarrow
  (\hat{\bar\Psi}_-,\hat{\Upsilon}_+) = 
    ({\bar\Psi}_-,{\Upsilon}_+) + \i (D_-\bar\varphi, D_+\chi)
 \end{split} \label{e:PsiYtrans}
\end{gather}
and with $B_{A'B}\equiv-B_{BA'}$, the action \eqref{e:N=2.2action} becomes
\begin{gather}
    S= -\int \d^2\xi \d^2\theta \Big[
      \hat\Psi^{A}_{-} B_{AB^\prime}\hat\Upsilon^{B^\prime}_+
      +D_+\varphi^A B_{AB^\prime}D_-\chi^{B^\prime} 
      +D_+\chi^{A^\prime} B_{A^\prime B}D_-\varphi^{B}\Big].
  \label{e:N2.2toN1.1action}
\end{gather}
We find it convenient to collect the fields into
\begin{align}
  \phi^\alpha = (\varphi^A, \chi^{A^\prime}) &&
  \Psi^{\alpha}_{+} = (\hat\Psi^{A}_+, \hat\Upsilon^{A^\prime}_+) &&
  \Psi^\alpha_{-} = (\hat\Psi^A_{-}, \hat\Upsilon^{A^\prime}_-) 
\end{align}
and introduce
\begin{align}
  B_{\alpha\beta} = \matrix{cc}{
     0&B_{A B^\prime} \\ B_{A^\prime B} &0}.
\end{align}
We let $\S_{+\alpha} = \Psi^\kappa_+B_{\kappa\alpha}$ and
$\S_{-\alpha} = B_{\alpha\kappa}\Psi^\kappa_-$ and denote the inverse of
$B_{\alpha\beta}$ by $\Pi^{\alpha\beta}$.
We may then rewrite \eqref{e:N=2.2action} as
\begin{gather}
  S = - \int \d^2\xi \d^2\theta
    \Big[ \S_{+\alpha}\Pi^{\alpha\beta}\S_{-\beta} + 
          D_+\phi^\alpha B_{\alpha\beta}D_-\phi^\beta \Big]
\end{gather}
where $\Pi$ and $B$ are constant antisymmetric tensors by definition. This
implies that the second term vanishes, however, we keep it for clarity. Even
though $\Pi = B^{-1}$, we distinguish them to keep as close as
possible to the discussion in the previous sections. The fields $S_\pm$ are
constrained by 
\begin{align}
  \S_{-A} &= \S_{+A^\prime} = 0  &
  \hat\Upsilon^{A^\prime}_- &= \hat\Psi^{A}_+ = 0. \label{e:SFEQ}
\end{align}
The $2p+2p^\prime$ constraints on the $N=(2,2)$ fields \eqref{e:XYdef}
translate into restrictions on the auxiliary fields $\S_\pm$. Effectively, half
of them have been integrated out by means of their field equations
\eqref{e:SFEQ}. This is the direct translation of the
constraints \eqref{e:XYdef} on the $N=(2,2)$ fields. We may formally introduce
\begin{align}
  \Psi_+ = Q_+ \chiral{X}|&& \Upsilon_- = Q_- \chiral{Y}|&&
  \bar\Psi_+ = Q_+ \bar{\chiral{X}}|&&\bar\Upsilon_- = Q_- \bar{\chiral{Y}}|.
\end{align}
The constraints \eqref{e:XYdef} transform into
\begin{align}
  \Psi_+ = \i D_+ \varphi && \Upsilon_- = - \i D_- \chi &&
  \bar\Psi_+ = -\i D_+ \bar\varphi && \bar\Upsilon_- = \i D_- \bar\chi.
  \label{e:PsiYrest}
\end{align}
If we define the transformation \eqref{e:PsiYtrans} on these fields by 
\begin{gather}
 \begin{split}
  (\Psi_+,\bar\Upsilon_-) \rightarrow (\hat\Psi_+,\hat{\bar\Upsilon}_-) = 
  (\Psi_+,\bar\Upsilon_-) - \i (D_+\varphi,D_-\bar\chi) \\
  (\bar\Psi_+,\Upsilon_-) \rightarrow (\hat{\bar\Psi}_+,\hat\Upsilon_-) = 
  (\bar\Psi_+,\Upsilon_-) + \i (D_+\bar\varphi,D_-\chi)
 \end{split}
\end{gather}
we find that \eqref{e:SFEQ} and \eqref{e:PsiYrest} match each other. Thus,
using the field equations for half of the auxiliary $N=(1,1)$ fields is
equivalent to constraining the $N=(2,2)$ superfields.  By a simple rotation we
see that we cover all cases which allow for Darboux-Nijenhuis coordinates, even
for non-constant $B$.

By integrating out some of the fields, the almost complex structure matrices
effectively collapse to generalized complex structures: \begin{gather}
 \begin{split}
   \gen{J}^\p = \genMatrix{
     \ph{-}J^\p&-P^\p&\ph{-}0\\
     \ph{-}0&\ph{-}K^\p&\ph{-}0\\
     \ph{-}0&-Z^\p&\ph{-}R^\p
   } \longrightarrow \gcg{J}^\p =
   \gcgMatrix{
     \ph{-}\hat J^\p&-\hat P^\p \\
     \ph{-}0&\ph{-}\hat K^\p
   }.
 \end{split}
\end{gather}
In terms of $A,A^\prime$ coordinates, this reads
\begin{gather}
  \gcg{J}^\p = \matrix{cc|cc}{
    \ph{-}J^\p&\ph{-}0&\ph{-}0&\ph{-}0 \\
    \ph{-}0&\ph{-}J^{\p\prime}&-P^\p&\ph{-}0 \\
    \hline
    \ph{-}0&\ph{-}0&\ph{-}K^\p&\ph{-}0 \\
    \ph{-}0&\ph{-}0&-Z^\p&\ph{-}R^{\p\prime}
  } \label{e:JGCG}
\end{gather}
where we identified the tensors with their remaining components, e.g.\
$K^\alpha_\beta \rightarrow K^A_B$.  There is a similar reduction for
$\gen{J}^\m \rightarrow \gcg{J}^\m$.  Comparing to the results of
\cite{Lindstrom:2004hi}, we find their solution to match ours, \eqref{e:JGCG} if
$Z^\ppm=0$. We find $P^\ppm=-P^{\pmp t}$. It follows $J^\p = -J^\m$ and 
$R^\ppm = -K^\ppm$.  $Z^\ppm=0$ implies $[J^\ppm, \omega] = 0$, with $\omega$ as
defined in section \ref{s:solution} where we identified this case with a
symplectic manifold.

In the non-manifest description, we found a whole set of solutions given in
terms of almost complex structure matrices to choose among. In the left-/right-chiral
description, we start with `diagonal' objects, i.e.\ Darboux-Nijenhuis
coordinates. We learn that the different alternatives should collapse into one
and the same in (reduced) Darboux-Nijenhuis coordinates. This implies that we
can choose $Z^\ppm =0$ in
these cases. 

If we integrate out the remaining spinorial fields we reduce the geometry to the
ordinary case of two (ordinary, commuting) complex structures $J^\ppm$ acting on
$TM$.  We obtain the following diagram:
\begin{gather*}
  \gen{J}^\ppm \quad \xrightarrow{\S_{+A^\prime},\S_{-A}=0} \quad
  \gcg{J}^\ppm \quad \xrightarrow{\S_{+A},\S_{-A^\prime}=0} \quad
  J^\ppm
\end{gather*}

We find the line of argumentation valid even when replacing $B_{AB^\prime}$ by
a non-constant $E=G+B$ in \eqref{e:N=2.2action}, following the lines of
\cite{Lindstrom:2004hi}. This strongly suggests that this is the general way to
understand and embed left-/right-(anti-)chiral $N=(2,2)$ theories in this context.

\section{Discussion}
We presented a new framework that might lead to new insights on the way towards
a complete understanding of the geometry underlying $N=(2,2)$ supersymmetric
non-linear sigma models. As examples we considered the symplectic sigma model and
showed how the manifest theories in terms of left-/right-chiral superfields can be
interpreted and embedded into our new framework.

It is an open question, how to treat arbitrary first order sigma models in this
context. Most intriguing is the question what the proper integrability
conditions are, mainly due to the lack of a proper language. One might expect
integrability to work out in a similar way as Courant integrability generalizes
ordinary integrability. For the particular models we studied, we found a
description that is given by a covariantly constancy condition of the almost
complex structure matrices. We argued that this gives the integrability
condition for these models. The solution is based on the invertibility of one of
the complex structure tensors. We elaborated on possible generalizations of our
solution and compared them to the corresponding second order sigma model. 
The most general target space geometry allowing for
the extension of a general $N=(1,1)$ sigma model to $N=(2,2)$ supersymmetry is
still to be discovered, although we believe that that framework presented here
contributes an important step in that direction. 
 
\subsection*{Acknowledgments}
The authors thank M.~Zabzine and L.~Wulff for enlightening discussions. We would
also like to thank L.~Bergamin for useful comments. The
research of UL is supported by EU grant (Superstring theory) MRTN-2004-512194
and VR grant 621-2003-3454.


\begingroup\raggedright\endgroup

\end{document}